# Physics-informed neural networks for shock capturing in inviscid flows around an airfoil

Jiahao Song[1,2,3], Wenbo Cao[4,5], Weiwei Zhang[1,2,3,*]

[1] School of Aeronautics, Northwestern Polytechnical University, Xi'an 710072, China

[2] International Joint Institute of Artificial Intelligence on Fluid Mechanics, Northwestern Polytechnical University, Xi'an, 710072, China

[3] National Key Laboratory of Aircraft Configuration Design, Xi'an 710072, China

[4] Institute of AI for Industries, Chinese Academy of Sciences, Nanjing 211135, China

[5] Institute of Computing Technology, Chinese Academy of Sciences, Beijing 100190, China

* Corresponding author. E-mail: aeroelastic@nwpu.edu.cn

**Abstract** Physics-informed neural networks (PINNs) have shown promise in solving partial differential equations (PDEs). However, shock capturing remains challenging for PINNs, particularly in steady flows around an airfoil, where shocks may not even be captured. In this study, we point out that the steady Euler equations used to construct the loss function impose weak constraints, which make it difficult to correct the continuous function approximation preference of neural networks, causing gradient descent to converge to a smooth local optimum. Based on this insight, we propose to strengthen the physical constraints by reconstructing steady shock capturing as a temporal evolution process governed by the unsteady Euler equations. The temporal evolution law of the flow field guides gradient descent, thereby alleviating its tendency to converge to smooth local optima and ensuring shock formation. We introduce a PDE loss function incorporating a pseudo-time stepping mechanism to obtain the steady state solution without solving the unsteady Euler equations over a long time interval. In addition, we develop a simplified formulation of the Euler equations to reduce the complexity of the PDE loss function, thereby improving shock capturing accuracy. By solving four forward problems involving different flow conditions and geometries, we verify that the proposed method achieves high accuracy in shock capturing while reducing the wall time to one-seventh of that required by existing advanced methods. Moreover, the proposed method demonstrates its capability to solve a parametric problem involving the angle of attack and Mach number without relying on any labeled data.

**Keywords** Shock, Physics-informed neural networks, Temporal evolution, Equation simplification, Airfoil

## 1. Introduction

Numerical solution of partial differential equations (PDEs) is one of the core problems in scientific computing. Traditional numerical methods, such as finite difference, finite volume, and finite element methods, commonly rely on discrete meshes, and their accuracy and applicability are largely affected by factors such as mesh quality and discretization schemes. Recently, with the rapid development of deep learning in function approximation and high-dimensional optimization, numerous methods based on neural networks have emerged for solving PDEs, among which physics-informed neural networks (PINNs) [1] are the most representative. The core idea of PINNs is to approximate solutions of PDEs by minimizing a composite loss function that penalizes PDE residuals and mismatches with the boundary and initial conditions. Compared with traditional numerical methods, PINNs have the advantages of being meshfree and enabling the solution of inverse problems [2-4] and parametric problems [5, 6]. The method has been applied in various scientific fields, including fluid mechanics [7-9], materials [10, 11], geosciences [12, 13], and bio-engineering [14, 15].

In fluid mechanics, shocks are highly nonlinear flow phenomena characterized by intense local variations caused by compressibility effects. When the upstream Mach number normal to the shock exceeds unity, small pressure disturbances cannot propagate upstream against the incoming flow, and the flow properties change abruptly across the shock. Strong local gradients make shock capturing one of the major challenges in computational fluid dynamics (CFD). Several studies have applied PINNs to capture shocks in fluid mechanics. Mao et al. [16] solved inviscid supersonic problems using PINNs, and they improved the performance of the algorithm by densifying collocation points near the shocks. By introducing observed data into PINNs, Jagtap et al. [17] solved inverse problems involving shocks. Patel et al. [18] constructed the loss function by incorporating ideas from the finite volume method, such as the entropy condition and the total variation diminishing constraint, thereby improving the applicability of PINNs to shock capturing. Liu et al. [19] assigned weights to the PDE residuals at each collocation point in the loss function based on the velocity gradient, thereby preventing high PDE residuals caused by large flow gradients near shocks from dominating gradient descent. However, these studies mainly deal with problems such as shock tubes and oblique shocks, while flows around an airfoil, which are directly relevant to aerodynamic analysis and design, have received limited attention. Cao et al. [20] proposed NNfoil, a solver for steady inviscid subsonic flows around an airfoil, by combining PINNs with mesh transformation. The nonuniform collocation points in the physical space are transformed into uniform collocation points in the computational space, thereby significantly improving the accuracy of PINNs. They also showed that NNfoil could not capture shocks under transonic inflow conditions. Based on this method, Wassing et al. [21] solved inviscid transonic flows around an airfoil by introducing artificial viscosity into the steady Euler equations. However, the results presented in their study still show noticeable errors near the shocks. Recently, based on [21], Lei et al. [22] introduced the Kolmogorov-Arnold Network (KAN), achieving more accurate shock capturing in inviscid flows around an airfoil. However, their method has a high computational cost, requiring 9.58 hours on an NVIDIA GeForce RTX 4090 GPU. Therefore, this study aims to develop a highly accurate and efficient shock capturing method for PINNs in steady inviscid flows around an airfoil.

We propose a key insight: for steady inviscid flows around an airfoil, neural networks are capable of approximating shocks. However, in NNfoil, the steady Euler equations used to construct the PDE loss function impose weak constraints, making it difficult to correct the continuous function approximation preference [23-25] of neural networks. As a result, NNfoil tends to converge to a smooth local optimum when directly approximating steady flow fields with shocks. Accordingly, we propose to reconstruct steady shock capturing as a process of temporal evolution that gradually converges to the steady state solution. NNfoil is constrained by the unsteady Euler equations. Gradient descent is guided by the temporal evolution law of the flow field. This additional temporal penalty strengthens the physical constraint, thereby alleviating the tendency of NNfoil to converge to a smooth local optimum. Considering that the unsteady Euler equations need to be solved over a long time to obtain the steady state solution and that the capability of PINN-like methods to solve long time problems is poor [26-28], we introduce a PDE loss function [29] that embeds the concept of pseudo-time stepping to avoid this issue. In addition, we develop a simplified formulation of the Euler equations to reduce the complexity of the PDE loss function, thereby improving shock capturing accuracy.

The remainder of the paper is organized as follows. In Section 2, we provide a brief introduction to PINNs and NNfoil, followed by a detailed presentation of the proposed method. Other techniques used in this study are also reviewed. In Section 3, we validate the effectiveness of the proposed method by solving four forward problems with different flow conditions and geometries, as well as a parametric problem involving the angle of attack and Mach number. Finally, Section 4 provides concluding remarks and suggests directions for future research.

## 2. Methodology

*2.1 Physics-informed neural networks*

PINNs transform the solution of PDEs into an optimization problem that aims to minimize the loss function. They take the space-time coordinates $(\boldsymbol{x}, t)$ as the inputs, and output the approximate solution $\boldsymbol{U}$. We consider the dimensionless steady two-dimensional Euler equations defined on a domain $\Omega \subset \mathbb{R}^2$:

$$\frac{\partial \boldsymbol{F}}{\partial x} + \frac{\partial \boldsymbol{G}}{\partial y} = 0$$

$$\boldsymbol{F} = \begin{Bmatrix} \rho u \\ \rho u^2 + p \\ \rho uv \\ u(E+p) \end{Bmatrix}, \quad \boldsymbol{G} = \begin{Bmatrix} \rho v \\ \rho uv \\ \rho v^2 + p \\ v(E+p) \end{Bmatrix} \tag{1}$$

where the first equation is the continuity equation, the second and third equations are the momentum equations in the $x$ and $y$ directions, and the fourth equation is the energy equation. $\rho$ represents the density. $u$ and $v$ are the $x$ and $y$ components of the velocity vector $\boldsymbol{V}$. $p$ represents the pressure. $E$ is the total energy per unit volume. They satisfy the following relation

$$E = \frac{p}{\gamma - 1} + \frac{1}{2}\rho(u^2 + v^2) \tag{2}$$

where $\gamma = 1.4$ is the specific heat ratio. Based on Eq. (2), Eq. (1) can be transformed into a closed system of equations with respect to $\boldsymbol{U} = [\rho, \rho u, \rho v, E]$. The boundary conditions are given by $\mathcal{B}[\boldsymbol{U}] = 0$.

The loss function of PINNs is

$$\mathcal{L} = w_{bc}\mathcal{L}_{bc} + w_r\mathcal{L}_r \tag{3}$$

where

$$\mathcal{L}_{bc} = \frac{1}{N_{bc}} \sum_{i=1}^{N_{bc}} \left| \mathcal{B}[\boldsymbol{U}_\theta(x_{bc}^i, y_{bc}^i)] \right|^2 \tag{4}$$

$$\mathcal{L}_r = \frac{1}{N_r}\sum_{i=1}^{N_r}\left|\frac{\partial \boldsymbol{F}[\boldsymbol{U}_\theta(x_r^i, y_r^i)]}{\partial x} + \frac{\partial \boldsymbol{G}[\boldsymbol{U}_\theta(x_r^i, y_r^i)]}{\partial y}\right|^2 \tag{5}$$

in Eqs. (4)-(5), $\theta$ represent the network parameters. $\{x_{bc}^i, y_{bc}^i\}_{i=1}^{N_{bc}}$ represent training points for the boundary conditions, and the boundary condition loss $\mathcal{L}_{bc}$ constrains the outputs of the PINNs to satisfy the boundary conditions. $\{x_r^i, y_r^i\}_{i=1}^{N_r}$ represent collocation points, and the PDE loss $\mathcal{L}_r$ ensures that the outputs comply with the PDEs. $w_{bc}$ and $w_r$ are weight factors used to balance the two losses, thereby ensuring that both are sufficiently reduced. For unsteady problems, Eq. (3) further includes the initial condition loss $\mathcal{L}_{ic}$ and the corresponding weight factor $w_{ic}$.

Cao et al. [20] combined PINNs with mesh transformation to improve the performance of PINNs in solving steady inviscid subsonic flows around an airfoil, and named the proposed method NNfoil. Mesh transformation is a method that maps the physical space $(x, y)$ to the computational space $(\xi, \eta)$ through the mappings $\xi = \xi(x, y)$ and $\eta = \eta(x, y)$. In the computational space, the flow field region near the airfoil is enlarged, thereby reducing the difficulty for PINNs to approximate the flow field, as shown in Fig. 1.

In [20], the effectiveness of NNfoil was validated through extensive numerical experiments, while its limitation in capturing shocks was also demonstrated. NNfoil is employed in this study. After mesh transformation, the network input becomes $(\xi, \eta)$, so $\partial \boldsymbol{F} / \partial \xi$, $\partial \boldsymbol{F} / \partial \eta$, $\partial \boldsymbol{F} / \partial \eta$ and $\partial \boldsymbol{G} / \partial \eta$ can be directly obtained through automatic differentiation [30]. Furthermore, the partial derivatives $\partial \boldsymbol{F} / \partial x$ and $\partial \boldsymbol{G} / \partial y$ in the physical space are calculated using Eq. (6).

$$\begin{aligned} \frac{\partial \boldsymbol{F}}{\partial x} &= \frac{\partial \boldsymbol{F}}{\partial \xi}\frac{\partial \xi}{\partial x} + \frac{\partial \boldsymbol{F}}{\partial \eta}\frac{\partial \eta}{\partial x} \\ \frac{\partial \boldsymbol{G}}{\partial y} &= \frac{\partial \boldsymbol{G}}{\partial \xi}\frac{\partial \xi}{\partial y} + \frac{\partial \boldsymbol{G}}{\partial \eta}\frac{\partial \eta}{\partial y} \end{aligned} \tag{6}$$

where $\partial \xi / \partial x$, $\partial \xi / \partial y$, $\partial \eta / \partial x$ and $\partial \eta / \partial y$ are metrics, which are obtained from the mappings $\xi = \xi(x, y)$ and $\eta = \eta(x, y)$. Since the mappings in this study are not analytical, the metrics are obtained using second-order finite differences. The discontinuous boundaries $\Gamma_3$ and $\Gamma_4$ in the computational space correspond to the same location in the physical space. Therefore, periodic boundary conditions need to be imposed at $\Gamma_3$ and $\Gamma_4$ in addition to the original boundary conditions.

In addition, the volume weighting PDE loss function (7) proposed by Song et al.

[31] is employed in NNfoil to replace the standard PDE loss (5). In this study, we adopt the same setting. In Eq. (7), $s(x_r^i, y_r^i)$ represents the volume occupied by collocation point $(x_r^i, y_r^i)$ in the physical space. Since the volume weighting PDE loss is generally several orders of magnitude lower than the boundary condition loss, a weight $a = 2\times10^4$, independent of $w_r$, is introduced to ensure that both losses are sufficiently reduced during gradient descent.

$$\mathcal{L}_r = \frac{1}{N_r} a \sum_{i=1}^{N_r} \left| \left( \frac{\partial \boldsymbol{F}[\boldsymbol{U}_\theta(x_r^i, y_r^i)]}{\partial x} + \frac{\partial \boldsymbol{G}[\boldsymbol{U}_\theta(x_r^i, y_r^i)]}{\partial y} \right) s(x_r^i, y_r^i) \right|^2 / \sum_{i=1}^{N_r} [s(x_r^i, y_r^i)]^2 \tag{7}$$

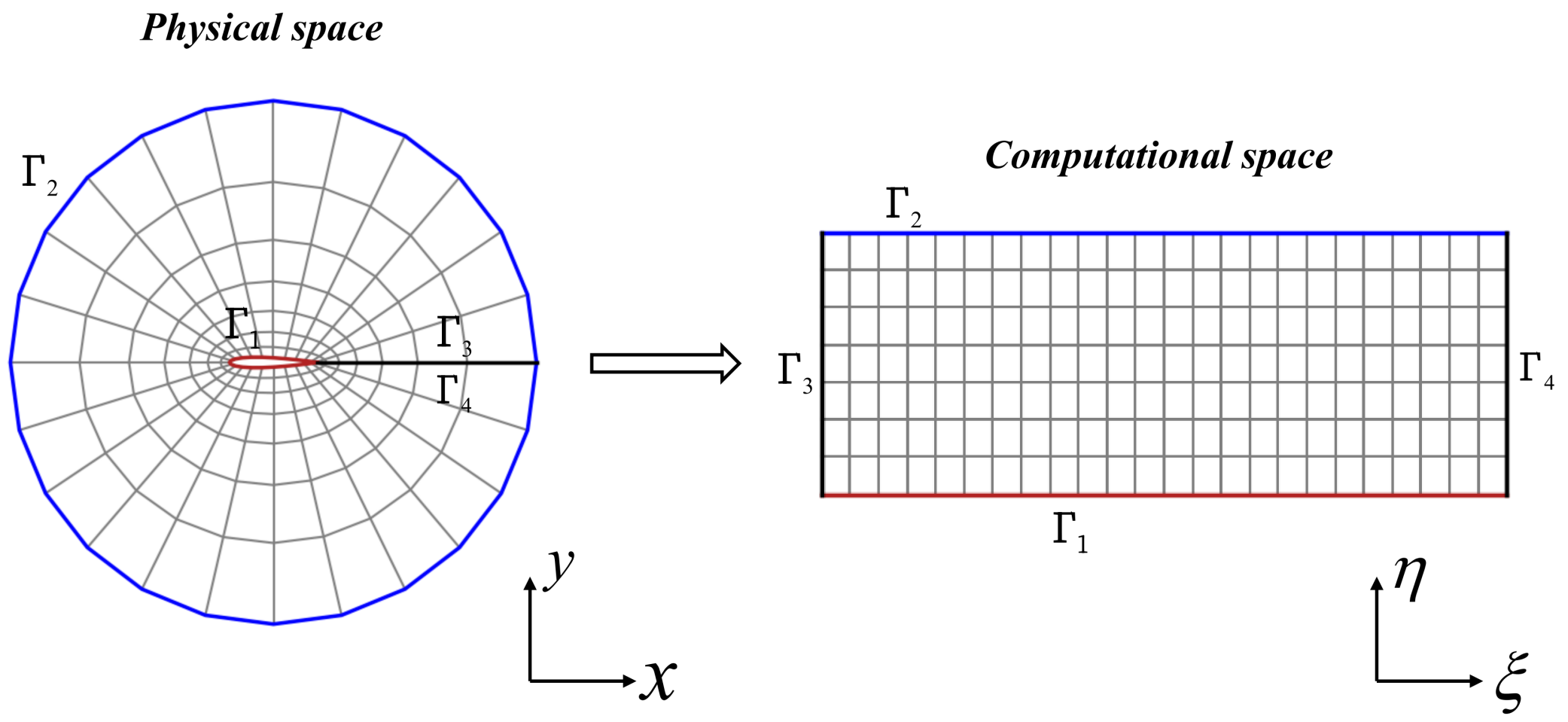

Figure 1. A schematic of mesh transformation for solving partial differential equations.

*2.2 Down-weighting of PDE residuals near shocks based on velocity gradient*

In [19], Liu et al. pointed out that the steep gradients of flow variables near shocks lead to much larger PDE residuals there than in other regions. Consequently, the gradient descent direction is dominated by the residuals in this region, while other regions receive relatively less attention, making it difficult for PINNs to obtain satisfactory results. To address this issue, the authors designed a weight $\lambda_r$ based on the velocity gradient for the PDE residuals in the loss function.

$$\lambda_r(x_r^i, y_r^i) = \frac{1}{k(\left|\nabla\cdot\boldsymbol{V}(x_r^i, y_r^i)\right| - \nabla\cdot\boldsymbol{V}(x_r^i, y_r^i)) + 1} \tag{8}$$

where $k > 0$ is an empirical parameter. It can be observed that near shocks, $\left|\nabla\cdot\boldsymbol{V}\right| - \nabla\cdot\boldsymbol{V}$ is very large, and the corresponding $\lambda_r$ tends to 0. In other regions where the flow field is smooth, $\left|\nabla\cdot\boldsymbol{V}\right| - \nabla\cdot\boldsymbol{V}$ is small, and $\lambda_r$ tends to 1. Therefore, using $\lambda_r$ as the weight can prevent excessively large PDE residuals near shocks from

dominating gradient descent. The effectiveness of the method was validated by shock tube problems and two-dimensional Riemann problems in [19]. The method is also employed in this study. However, it should be noted that, for steady inviscid flows around an airfoil, NNfoil still fails to capture shocks when only this method is employed (the numerical results are presented in Section 3).

*2.3 Strengthening physical constraints through temporal evolution for shock capturing*

We first identify a key factor contributing to the failure of NNfoil to capture shocks: for steady inviscid flows around an airfoil, neural networks are capable of approximating shocks. However, in NNfoil, although the steady Euler equations used to construct the PDE loss function are sufficient from a physical perspective, they impose weak constraints in the optimization problem. A smooth solution may still yield small residuals of the steady Euler equations, making the loss function insensitive to its difference from the true solution with shocks. Consequently, the constraints imposed by the steady Euler equations make it difficult to correct the continuous function approximation preference [23-25] of neural networks, causing gradient descent to converge to a smooth local optimum.

Based on this insight, we propose to strengthen the physical constraints by reconstructing steady shock capturing as a temporal evolution process that gradually converges to the steady state solution. Specifically, the unsteady Euler equations (9) constructed by introducing time derivative terms into the steady equations are used to constrain NNfoil. The output of the neural network is no longer required to directly approximate a flow field with shocks by minimizing the residuals of the steady Euler equations. Instead, shocks gradually form under the guidance of the temporal evolution law of the flow. This additional temporal penalty strengthens the physical constraint to alleviate the tendency of NNfoil to converge to a smooth local optimum, thereby improving its ability to capture shocks. From the perspective of PDE classification, for most flow fields involving shocks, the steady Euler equations are generally of mixed type, with different information propagation properties in different regions of the flow field. In contrast, the unsteady Euler equations are hyperbolic throughout the entire physical domain, thereby establishing a unified mechanism for information propagation. We demonstrate the effectiveness of using the unsteady Euler equations to constrain NNfoil for shock capturing by solving the inviscid flow around the NACA0012 airfoil at Mach number $Ma = 0.7$ and angle of attack $\alpha = 5°$. Detailed results are presented in Appendix A.

$$\frac{\partial \boldsymbol{U}}{\partial t}+\frac{\partial \boldsymbol{F}}{\partial x}+\frac{\partial \boldsymbol{G}}{\partial y}=0$$

$$\boldsymbol{U}=\begin{Bmatrix}\rho \\ \rho u \\ \rho v \\ E\end{Bmatrix},\quad \boldsymbol{F}=\begin{Bmatrix}\rho u \\ \rho u^2+p \\ \rho uv \\ u(E+p)\end{Bmatrix},\quad \boldsymbol{G}=\begin{Bmatrix}\rho v \\ \rho uv \\ \rho v^2+p \\ v(E+p)\end{Bmatrix} \tag{9}$$

A key issue is that obtaining the steady state solution requires solving the unsteady equations over a long time in the time dimension. However, since the capability of PINN-like methods for solving long time problems is poor [26-28], directly solving the unsteady equations is impractical. To address this issue, we introduce the PDE loss function (10) proposed by Cao et al. [29], which embeds the concept of pseudo-time stepping.

$$\mathcal{L}_r=\frac{1}{N_r}\sum_{i=1}^{N_r}\left|\left(\frac{\partial \boldsymbol{F}[\boldsymbol{U}_\theta(x_r^i,y_r^i)]}{\partial x}+\frac{\partial \boldsymbol{G}[\boldsymbol{U}_\theta(x_r^i,y_r^i)]}{\partial y}\right)\Delta\tau+(\boldsymbol{U}_\theta(x_r^i,y_r^i)-\boldsymbol{U}_p(x_r^i,y_r^i))\right|^2 \tag{10}$$

where $\Delta\tau$ denotes the pseudo-time step. $\boldsymbol{U}_p$ denotes the solution output by NNfoil at the previous epoch, and it is detached from the computational graph. The term inside the absolute value operator $|\cdot|$ in Eq. (10) can be transformed into

$$\boldsymbol{U}_\theta(x_r^i,y_r^i)=\boldsymbol{U}_p(x_r^i,y_r^i)-\left(\frac{\partial \boldsymbol{F}[\boldsymbol{U}_\theta(x_r^i,y_r^i)]}{\partial x}+\frac{\partial \boldsymbol{G}[\boldsymbol{U}_\theta(x_r^i,y_r^i)]}{\partial y}\right)\Delta\tau \tag{11}$$

according to Eq. (9), $-(\partial \boldsymbol{F}[\boldsymbol{U}_\theta]/\partial x+\partial \boldsymbol{G}[\boldsymbol{U}_\theta]/\partial y)$ is equal to $\partial \boldsymbol{U}_\theta/\partial t$. Therefore, gradient descent based on the PDE loss function (10) incorporates a pseudo-time stepping mechanism for $\boldsymbol{U}_\theta$, and the steady state solution can be obtained after an appropriate number of epochs. Since Eq. (10) is constructed using the steady Euler equations, solving the unsteady Euler equations over a long time in the time dimension is avoided. In this study, following the established design of Cao et al. [29], we set $\Delta\tau=0.3$.

*2.4 Simplified formulation of the steady Euler equations*

In traditional numerical methods, the conservative steady Euler equations (1) are commonly used to solve inviscid flows. Their expanded form is given by

$$
\begin{aligned}
&(\rho u)_x + (\rho v)_y = 0 \\
&\rho u u_x + \rho v u_y + p_x + u[(\rho u)_x + (\rho v)_y] = 0 \\
&\rho u v_x + \rho v v_y + p_y + v[(\rho u)_x + (\rho v)_y] = 0 \\
&\frac{\gamma}{\gamma-1}(pu)_x + \frac{\gamma}{\gamma-1}(pv)_y + \rho u(\frac{V^2}{2})_x + \rho v(\frac{V^2}{2})_y + \frac{V^2}{2}[(\rho u)_x + (\rho v)_y] = 0
\end{aligned}
\tag{12}
$$

However, in PINN-like methods, more complex PDEs tend to make the training more difficult [26, 32, 33]. Thus, we remove the products of the continuity equation with $u$, $v$, and $V^2/2$ from the momentum and energy equations in Eq. (12), respectively, and construct a simpler form of the steady Euler equations (13) as the PDE constraint in NNfoil.

$$
\begin{aligned}
&(\rho u)_x + (\rho v)_y = 0 \\
&\rho u u_x + \rho v u_y + p_x = 0 \\
&\rho u v_x + \rho v v_y + p_y = 0 \\
&\frac{\gamma}{\gamma-1}(pu)_x + \frac{\gamma}{\gamma-1}(pv)_y + \rho u(\frac{V^2}{2})_x + \rho v(\frac{V^2}{2})_y = 0
\end{aligned}
\tag{13}
$$

In fact, using the momentum equations in the $x$ and $y$ directions, the energy equation in Eq. (13) can be further simplified into the following pressure form

$$
u p_x + v p_y + \gamma p(u_x + v_y) = 0 \tag{14}
$$

however, the numerical results presented in Appendix B show that using Eq. (13) to construct the PDE loss function yields lower errors than replacing its energy equation with Eq. (14). This may be related to the elimination of the kinetic energy transport terms in Eq. (14).

Overall, the PDE loss function constructed in this study is given by

$$
\begin{aligned}
&\mathcal{L}_r = \frac{1}{N_r}\sum_{i=1}^{N_r}\left|\lambda_r(x_r^i, y_r^i)[R(x_r^i, y_r^i)S(x_r^i, y_r^i)\Delta\tau + (\boldsymbol{U}_\theta(x_r^i, y_r^i) - \boldsymbol{U}_p(x_r^i, y_r^i))]\right|^2 \\
&S(x_r^i, y_r^i) = s(x_r^i, y_r^i) / \sqrt{\frac{1}{N_r}\sum_{i=1}^{N_r}[s(x_r^i, y_r^i)]^2}
\end{aligned}
\tag{15}
$$

where $R(x_r^i, y_r^i)$ denotes the PDE residual obtained by Eq. (13). The volume weighting method was originally used to modify the standard PDE loss defined by Eq. (5). However, the PDE loss used in this study is the loss function defined by Eq. (10), which embeds the concept of pseudo-time stepping. Therefore, the volume weighting form is adjusted accordingly. Specifically, in Eq. (15), the normalized volume $S(x_r^i, y_r^i)$ is combined with the pseudo-time step $\Delta\tau$ to form the local time step. Based on our experience, this weighting form is better than the standard volume weighting in this setting.

# 3. Results

We validate the effectiveness of the proposed method by solving four forward problems involving different flow conditions and geometries, as well as a parametric problem involving the angle of attack and Mach number. Since the proposed method consists of multiple components, ablation experiments are performed to sufficiently test the contribution of each component. Specifically, as shown in Table 1, NNfoil with only the volume weighting PDE loss is defined as Baseline1, and further introducing the down-weighting of PDE residuals near shocks is defined as Baseline2. Both are existing methods. On this basis, the introduction of the proposed temporal evolution method is defined as Algorithm1, and further incorporating the simplified formulation of the steady Euler equations is defined as Algorithm2. For all the cases, consistent with the training setup of Cao et al. [20], the fully connected neural network is constructed with 5 hidden layers, each containing 128 neurons. We use the tanh activation function and initialize the network parameters using a Gaussian distribution $\mathcal{N}(0,1)$. The limited-memory Broyden-Fletcher-Goldfarb-Shanno (L-BFGS) [34] optimizer is employed to perform gradient descent, with the maximum number of inner iterations per epoch set to 300. $w_{bc} = 2$, $w_r = 1$, and $k = 0.1$. In addition, although the solution vector of the Euler equations is usually defined as $\boldsymbol{U} = [\rho, \rho u, \rho v, E]$, following [20] and existing studies on shock capturing based on PINNs [16, 19, 21], we choose $[\rho, u, v, p]$ as the outputs of the neural network in this study. The reference solution is obtained using the finite volume method. The programs are developed based on the PyTorch platform and executed on an NVIDIA GeForce RTX 4090 GPU.

*3.1 Forward problems involving different flow conditions and airfoils*

We first solve forward problems involving different flow conditions and airfoils. The Mach numbers, angles of attack, and airfoils are listed in Table 2. NACA0012, NACA2418, RAE2822, and S2050 are selected as representative airfoils to cover a broad range of airfoil geometries, including symmetric, supercritical, and high-lift designs, as illustrated in Fig. 2. Cases 1–3 involve transonic conditions, and the corresponding flow fields contain shocks attached to the airfoil. Case4 involves a supersonic condition, and the shock in the corresponding flow field is a detached bow shock located in front of the airfoil.

Table 1. Components of different methods used in the ablation experiments. "√" indicates that the corresponding module is included.

| | NNfoil | Down-weighting | Temporal evolution | Simplification |
|---|---|---|---|---|
| Baseline1 | √ | | | |
| Baseline2 | √ | √ | | |
| Algorithm1 | √ | √ | √ | |
| Algorithm2 | √ | √ | √ | √ |

Table 2. Flow conditions and geometries for the four cases.

| | $Ma$ | $\alpha$ | Airfoil |
|---|---|---|---|
| Case1 | 0.7 | 5° | NACA0012 |
| Case2 | 0.75 | 3° | RAE2822 |
| Case3 | 0.8 | –4° | S2050 |
| Case4 | 1.5 | 0° | NACA2418 |

The physical domain and the distribution of collocation points for solving Case1 are shown in Fig. 3. For all four cases, the physical domain and the numbers of collocation points are the same, with differences only in the airfoils and the distributions of collocation points. The number of collocation points is $N_r = 16080$, corresponding to 80% of the 20100 collocation points in the entire physical domain, and these points are resampled at each epoch. Boundary conditions $\boldsymbol{V} \cdot \boldsymbol{n} = \mathbf{0}$ and $[\rho_\infty, u_\infty, v_\infty, p_\infty] = [1, \cos(\alpha), \sin(\alpha), 1/\gamma Ma^2]$ are imposed on the airfoil and the far-field boundary, respectively, with $N_{bc} = 201$ training points for each. Moreover, as described in Section 2.1, due to the use of mesh transformation, periodic boundary conditions need to be imposed, corresponding to the black line in the physical domain shown in Fig. 3, with $N_{bc} = 100$ training points. All methods listed in Table 1 are trained for 1000 epochs.

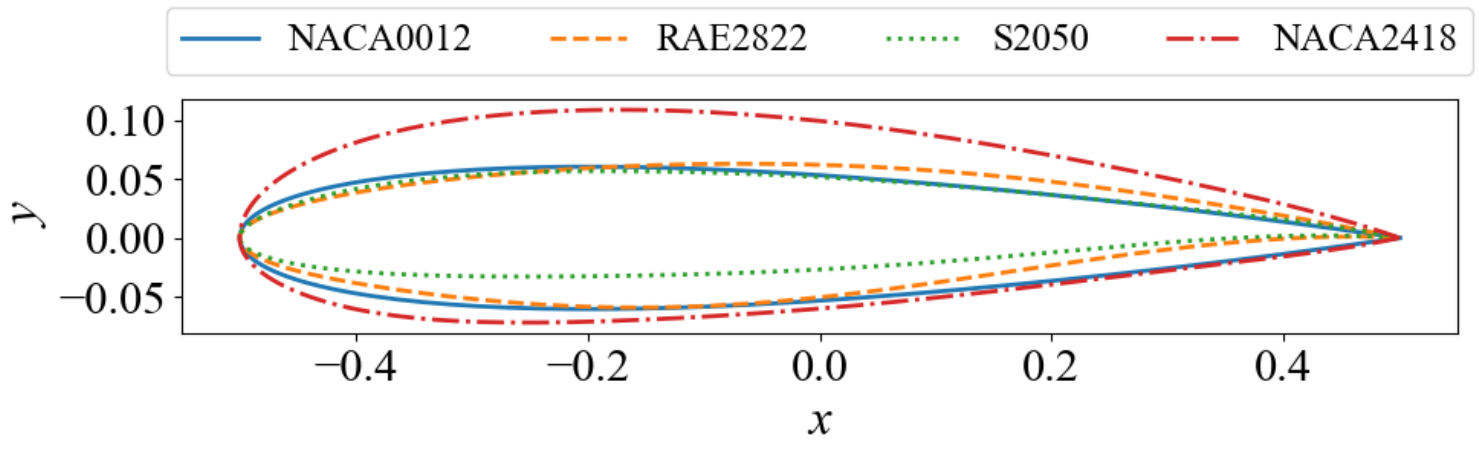

Figure 2. A schematic of four airfoils.

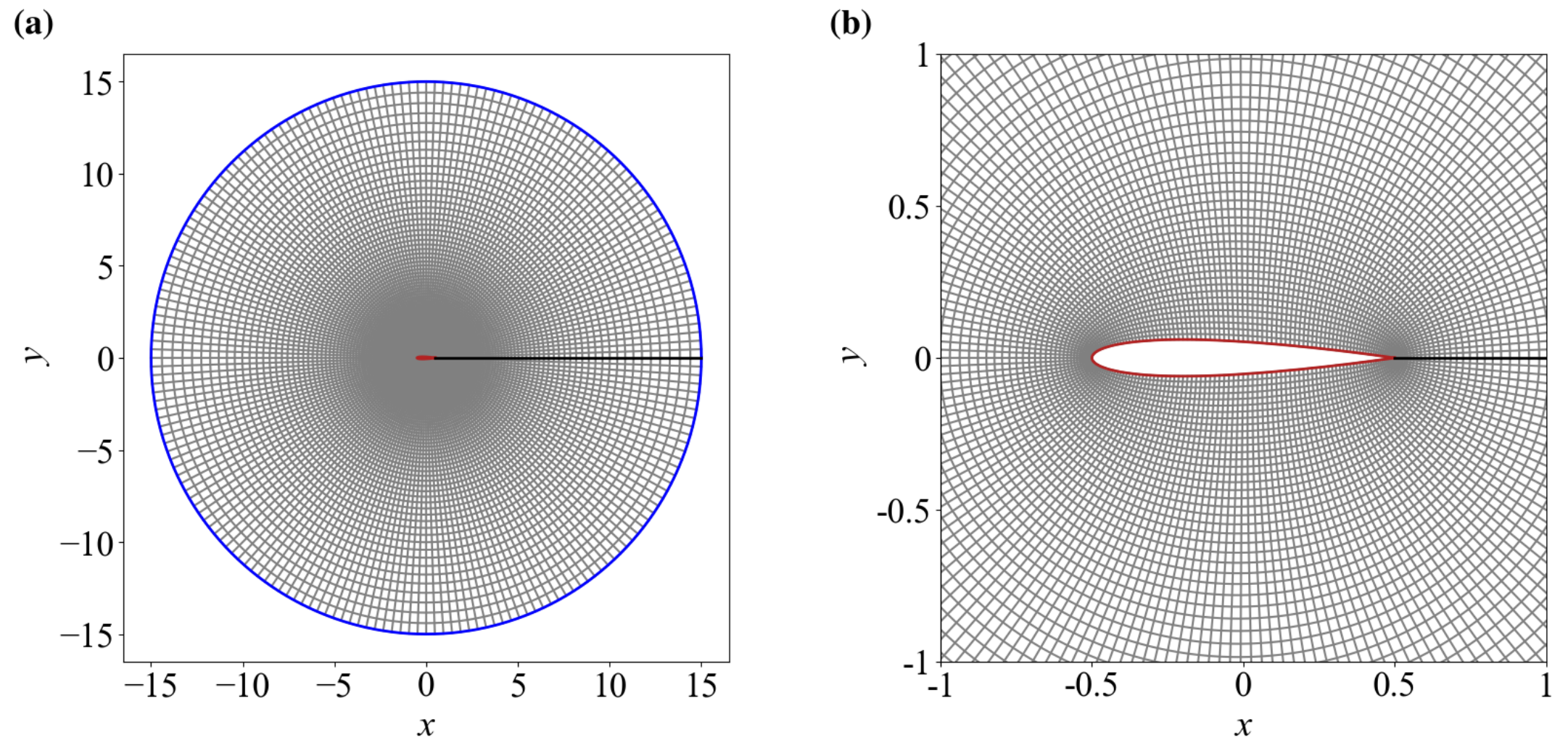

Figure 3. The physical domain and the distribution of collocation points for solving Case1. (a) Entire physical domain. The blue line represents the far-field boundary, the red line represents the NACA0012 airfoil, and the black line corresponds to the periodic boundary in the computational space. (b) Near the airfoil region.

The left and middle columns of Fig. 4 show the convergence histories of the boundary condition loss and PDE loss for the four cases. We evaluate the accuracy of the results using the pressure coefficient distribution $C_p = (p - p_\infty)/[0.5\rho_\infty(u_\infty^2 + v_\infty^2)]$ of the entire flow field. The convergence histories of the relative $L_2$ errors of $C_p$ are shown in the right column of Fig. 4. We observe that, although the final losses of all methods are similar, their relative $L_2$ errors differ significantly, as shown in Table 3. The errors of Baseline1 and Baseline2 generally exceed 30% and even approach 100% in the supersonic case because they converge to local optima. In contrast, the proposed Algorithm2 achieves errors below 3% in the transonic cases, while the error slightly increases to 7% in the supersonic case. Figure 5 shows the pointwise absolute errors of the four methods for Case1, and Fig. 6 shows those of Algorithm2 for the four cases. The results of the other three methods for Cases 2–4 are summarized in Appendix C. Furthermore, Fig. 7 shows the pressure coefficients on the airfoil obtained by the four methods for all cases. We observe that only Algorithm1 and Algorithm2 successfully capture the shocks, demonstrating the effectiveness of strengthening physical constraints through temporal evolution. Meanwhile, although Algorithm1 captures the shocks, deviations still exist in both shock strength and shock location. In contrast, the results obtained by Algorithm2 using the simplified Euler equations show good agreement with the reference solutions.

Table 4 summarizes the wall times required by the four methods. Although all methods are trained for the same number of epochs, differences exist in their wall times. The reason is that, in the L-BFGS optimizer, each epoch does not correspond to a fixed number of parameter updates. Instead, the optimizer adaptively determines the number of gradient computations required within each epoch according to the current iteration state, line search process, and convergence criteria. Therefore, the actual wall time depends not only on the prescribed number of epochs, but also on the optimization complexity of each method. The average wall time of Algorithm2 over the four cases is 1.42 hours. In comparison, Lei et al. [22] reported a wall time of 9.58 hours for their KAN-based method to achieve comparable accuracy using the same GPU (NVIDIA GeForce RTX 4090). Our method achieves a speedup of approximately 6.7.

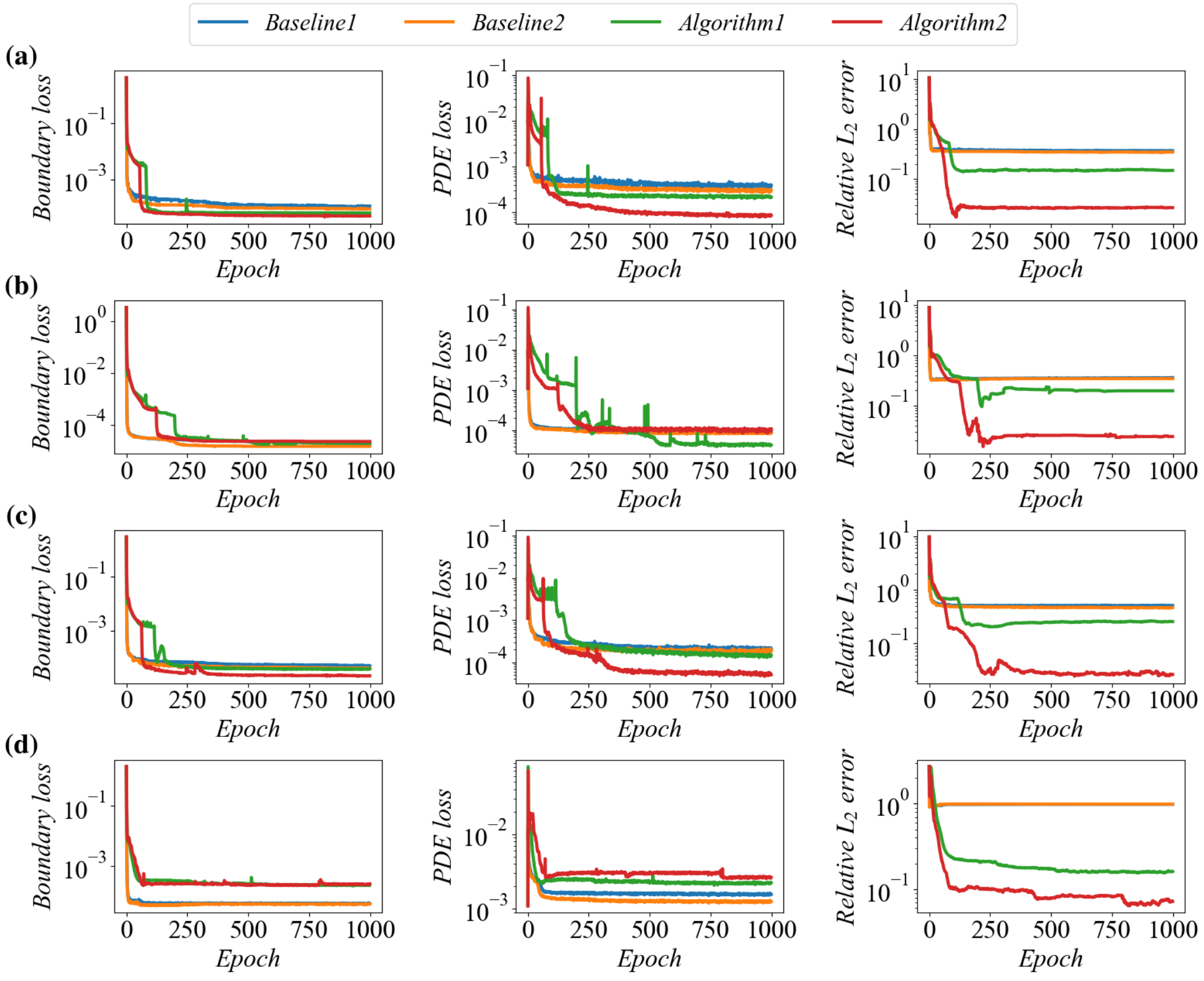

Figure 4. Convergence histories of different methods for solving forward problems involving different flow conditions and airfoils. (a) Case1. (b) Case2. (c) Case3. (d) Case4. For simplicity, the boundary condition loss is referred to as the boundary loss in the figure.

Taking Case1 as an example, we test the performance of Algorithm2 after removing each of its components individually, including NNfoil, Down-weighting, Temporal evolution, and Simplification, as shown in Fig. 8. We observe that the error near the shock increases significantly once any component is removed, indicating that all four components are essential for shock capturing.

Table 3. Comparison of relative $L_2$ errors (%) in the pressure coefficient distributions obtained by different methods for the four cases.

| | Baseline1 | Baseline2 | Algorithm1 | Algorithm2 |
|---|---|---|---|---|
| Case1 | 36.85 | 34.57 | 14.82 | **2.66** |
| Case2 | 35.87 | 34.40 | 19.66 | **2.41** |
| Case3 | 51.18 | 46.83 | 25.63 | **2.63** |
| Case4 | 98.48 | 98.98 | 16.10 | **7.20** |

In addition, since the results of neural networks exhibit randomness due to factors such as random initialization, we run Case1 five times for each method with different random seeds to evaluate the robustness. Detailed results are presented in Appendix D.

Table 4. Wall times of different methods for the four cases (unit: min).

| | Baseline1 | Baseline2 | Algorithm1 | Algorithm2 |
|---|---|---|---|---|
| Case1 | 66.82 | 65.58 | 74.49 | 71.87 |
| Case2 | 64.80 | 29.26 | 119.32 | 83.95 |
| Case3 | 80.97 | 66.70 | 100.33 | 85.16 |
| Case4 | 45.09 | 63.74 | 98.27 | 99.18 |

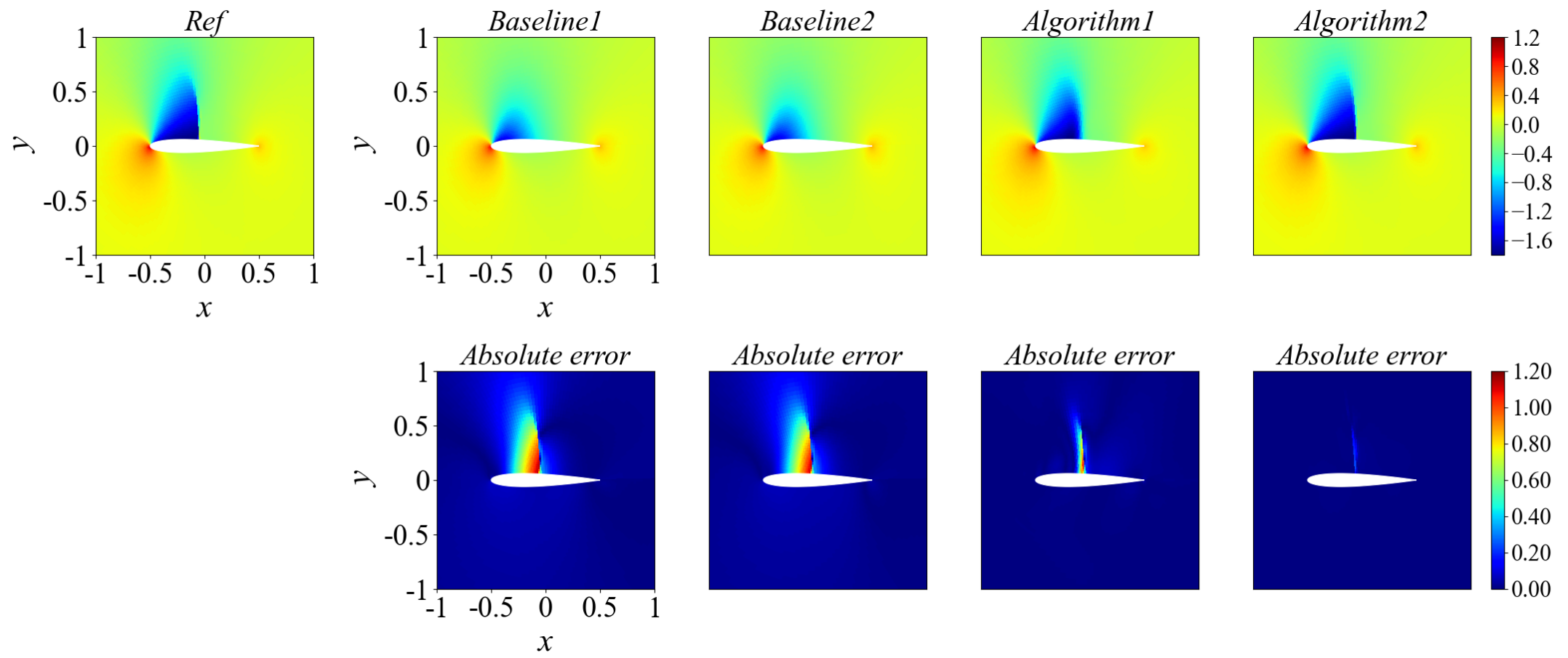

Figure 5. Comparison of pressure coefficient distributions obtained by different methods for Case1.

### *3.2 Parametric problem involving the angle of attack and Mach number*

Next, we solve a parametric problem involving the angle of attack and Mach number. PINNs can extend the independent variable space of the Euler equations from $(x, y)$ to include the angle of attack and Mach number as two additional dimensions by introducing these two parameters as network inputs, thereby obtaining all flows within the predefined angle of attack and Mach number parameter space in a single unsupervised training. In contrast, a single run of a traditional numerical method can only handle the flow at one angle of attack and Mach number. Therefore, for applications requiring flow field evaluations across multiple flow conditions and geometries, PINNs can significantly reduce the computational cost compared with traditional numerical methods [5, 35].

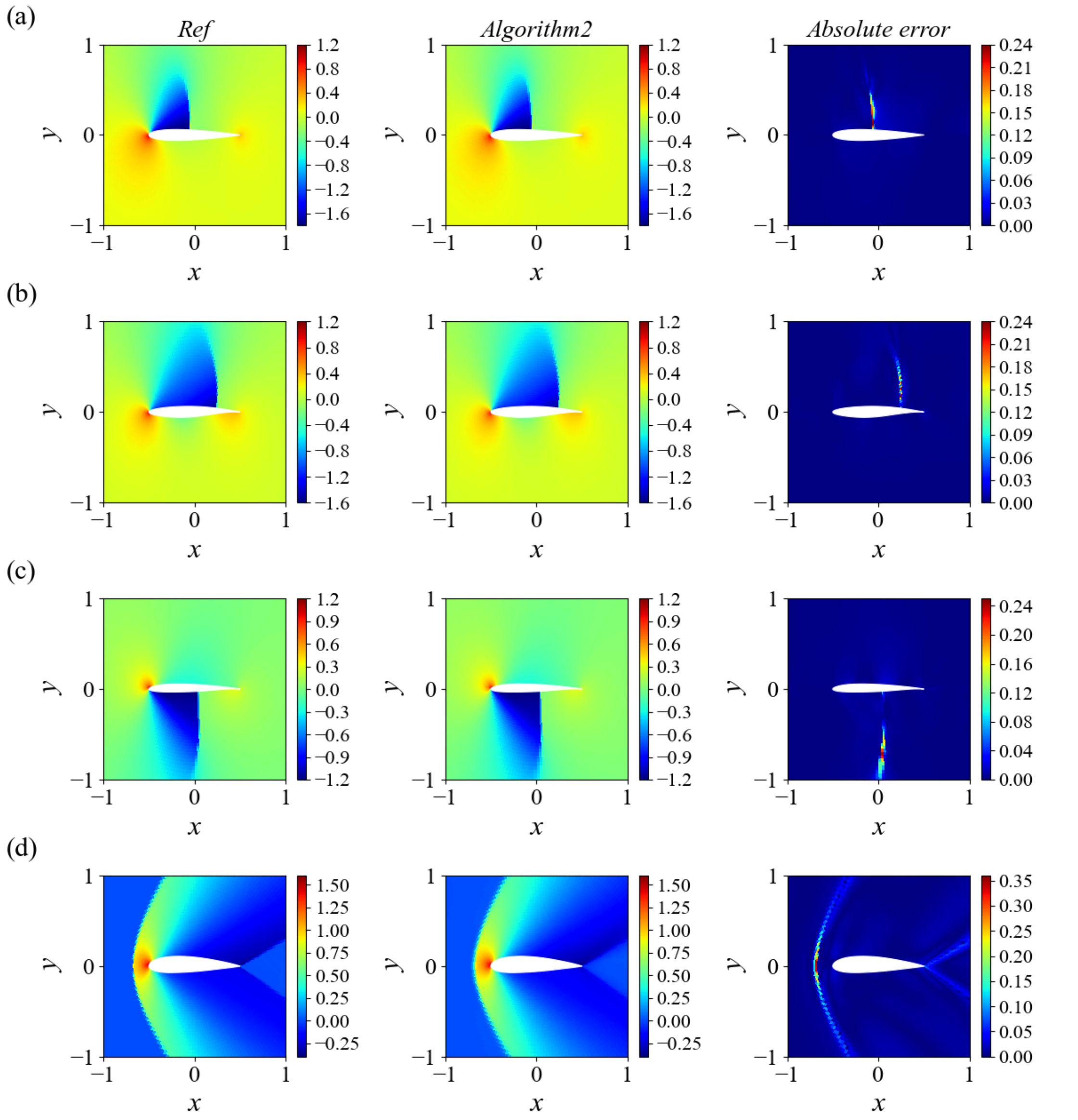

Figure 6. Comparison of pressure coefficient distributions obtained by Algorithm2 for the four cases. (a) Case1. (b) Case2. (c) Case3. (d) Case4.

We consider the NACA0012 airfoil, with $\alpha \in [-2^{\circ}, 7^{\circ}]$ and $Ma \in [0.5, 0.8]$. The physical domain, boundary conditions, and collocation point sampling strategy in the $(x, y)$ space are the same as those in Section 3.1. The number of collocation points is $N_r = 20000$, with resampling performed at each epoch. For the boundary conditions, $N_{bc} = 1005$ training points are used on each of the airfoil and far-field boundaries, and $N_{bc} = 500$ training points are used on the periodic boundary. Random sampling is used in the angle of attack and Mach number dimensions. Since only Algorithm2 achieves low error results for the forward problems, we only evaluate its performance in solving the parametric problem. The network is trained for 3000 epochs. The wall time is 7.77 hours.

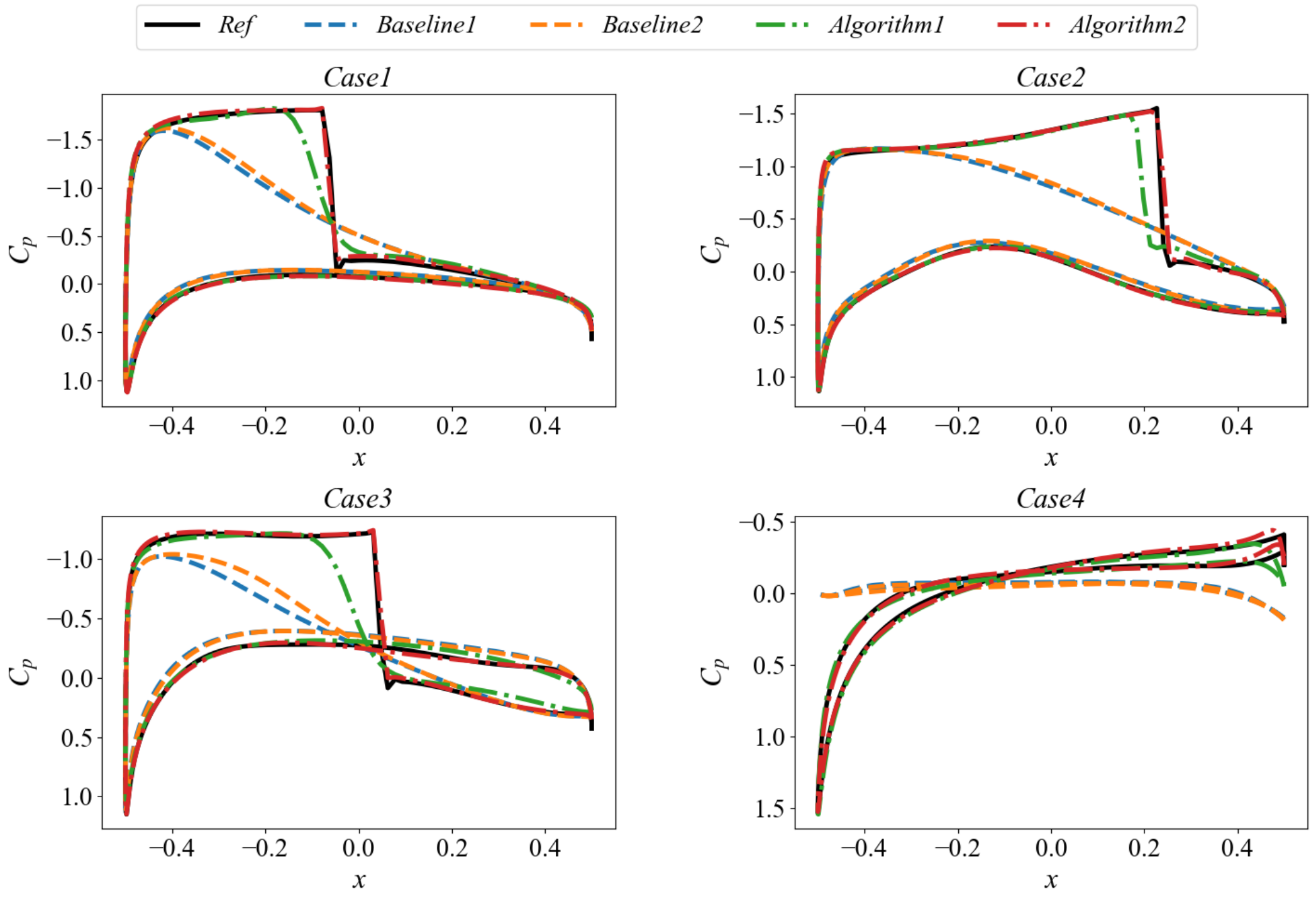

Figure 7. Comparison of pressure coefficients on the airfoil obtained by different methods for the four cases.

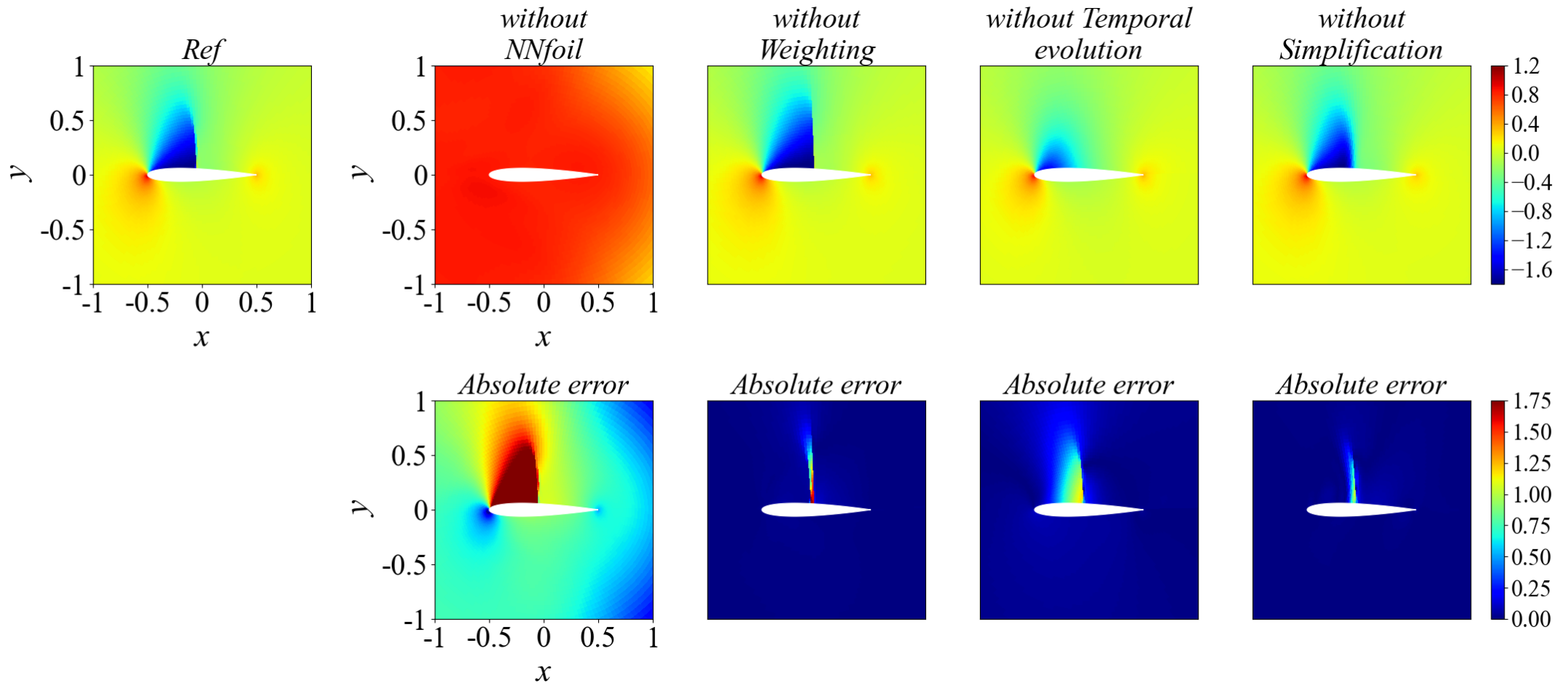

Figure 8. Comparison of pressure coefficient distributions obtained by the variants of Algorithm2 without NNfoil, Down-weighting, Temporal evolution, or Simplification for Case1.

Figure 9 shows the convergence histories of the loss functions for Algorithm2. Although the ranges of angle of attack and Mach number used for training are set to $\alpha \in [-2°, 7°]$ and $Ma \in [0.5, 0.8]$, the prediction ranges are restricted to $\alpha \in [0°, 5°]$ and $Ma \in [0.55, 0.75]$ because PINN-like methods have weak predictive capability near the boundaries of the parameter space [35]. Six different flow conditions are selected to evaluate the accuracy of Algorithm2, as listed in Table 5. Figure 10 compares the pressure coefficient distributions obtained by Algorithm2 with the reference solutions, and Fig. 11 further compares the pressure coefficients on the airfoil. We observe that the predicted results agree well with the reference solutions, demonstrating the effectiveness of Algorithm2 in solving the parametric problem. Although Zhou et al. [36] recently solved a similar problem, they introduced labeled data during training.

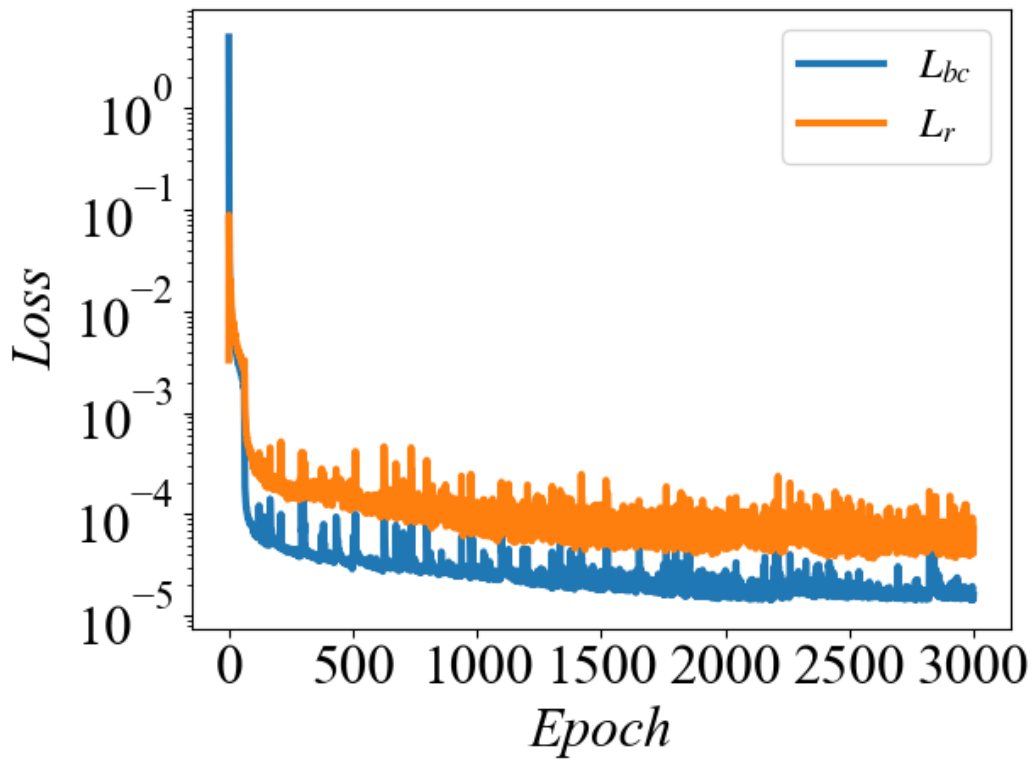

Figure 9. Convergence histories of the loss functions of Algorithm2 for solving the parametric problem.

Table 5. Six different flow conditions used to evaluate the accuracy of Algorithm2 for solving the parametric problem.

| Condition | 1 | 2 | 3 | 4 | 5 | 6 |
|---|---|---|---|---|---|---|
| $\alpha$ | 5° | 1° | 3° | 2° | 4° | 3° |
| $Ma$ | 0.55 | 0.6 | 0.65 | 0.7 | 0.7 | 0.75 |

## 4. Conclusions

In this study, we first identify the weak constraints imposed by the steady Euler equations as a key factor contributing to the failure of NNfoil to capture shocks. When the network outputs directly approximate a steady flow field with shocks, these constraints make it difficult to correct the continuous function approximation preference of neural networks, causing gradient descent to converge to a smooth local optimum.

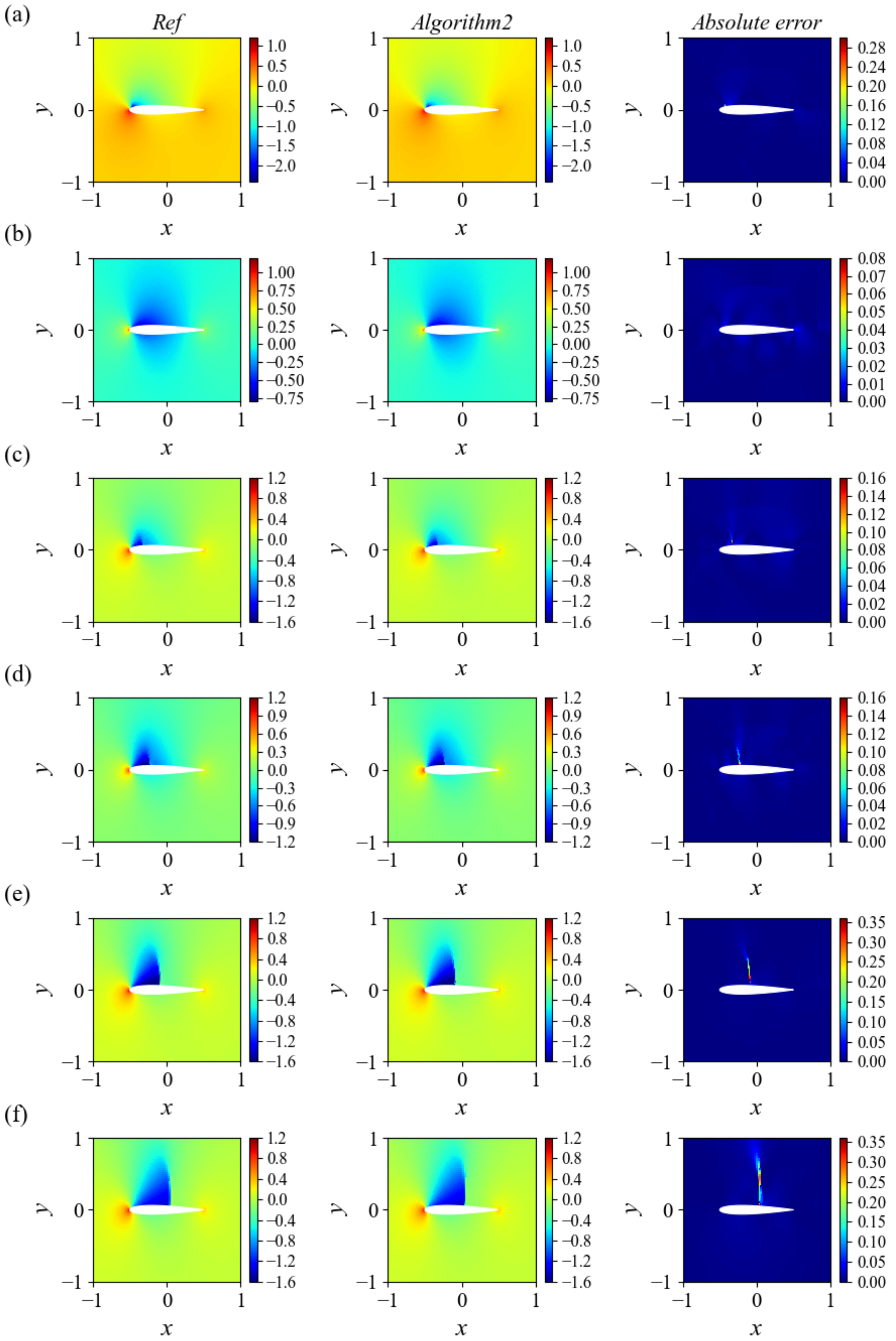

Figure 10. Pressure coefficient distributions at six different combinations of angle of attack and Mach number obtained by Algorithm2 for solving the parametric problem. The reference solutions are obtained from six independent runs of the finite volume method. (a) Condition1. (b) Condition2. (c) Condition3. (d) Condition4. (e) Condition5. (f) Condition6.

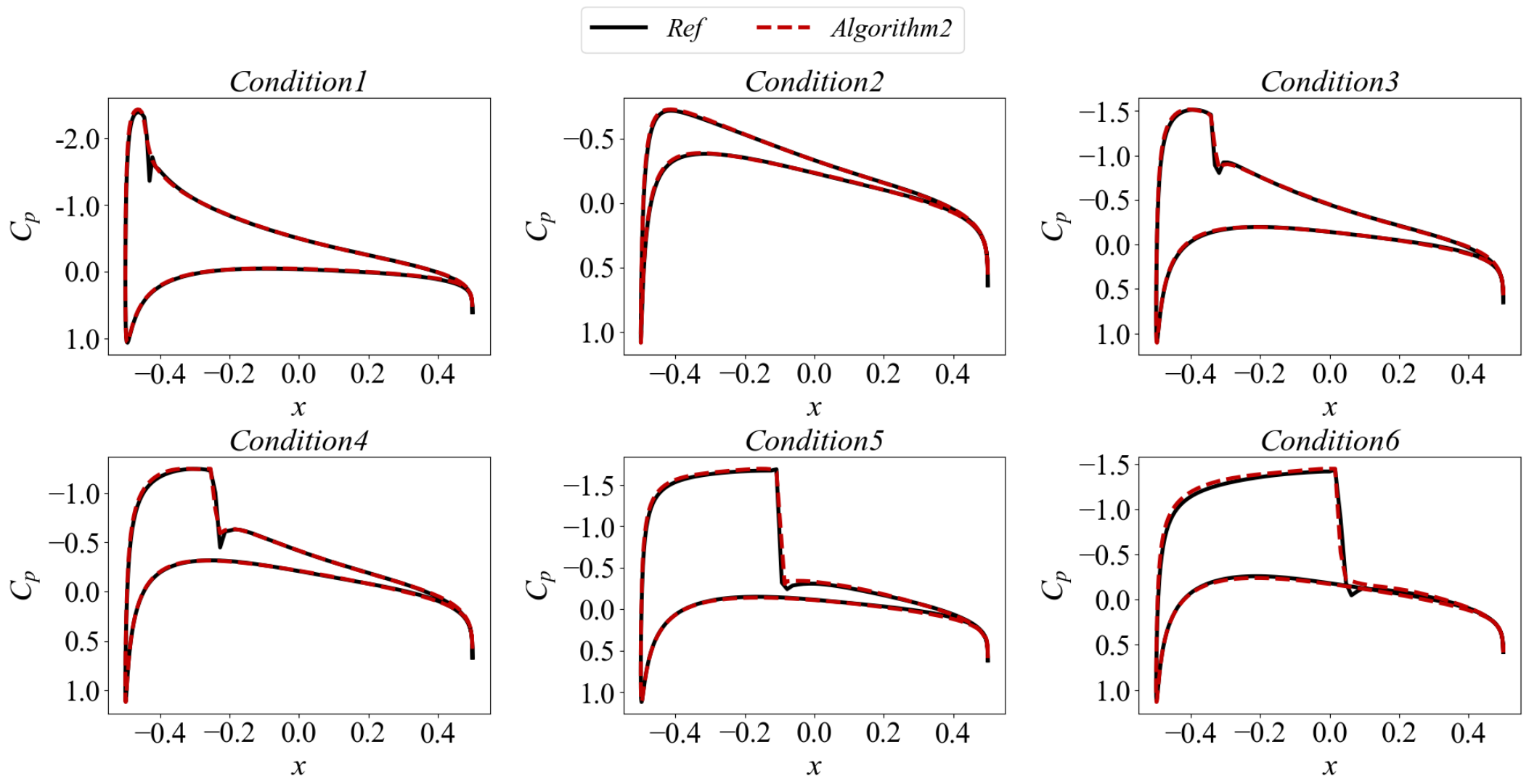

Figure 11. Pressure coefficients on the NACA0012 airfoil at six different combinations of angle of attack and Mach number obtained by Algorithm2 for solving the parametric problem. The reference solutions are obtained from six independent runs of the finite volume method.

Accordingly, we propose to reconstruct steady shock capturing as a process of temporal evolution that gradually converges to the steady state solution. By using the unsteady Euler equations instead of the steady equations as the PDE constraint of NNfoil, the network is required to follow the temporal evolution law of the flow to gradually form shocks. Compared with standard NNfoil, which directly approximates a flow field with shocks by minimizing the residuals of the steady Euler equations, our method imposes an additional temporal penalty on shock capturing to strengthen the physical constraints. The tendency of gradient descent to converge to a smooth local optimum is alleviated, ensuring that shocks form in the flow field. The capability of the proposed method is demonstrated by solving an unsteady problem. Since the unsteady Euler equations need to be solved over a long time in the time dimension to obtain the steady state solution, while the capability of PINN-like methods to solve such problems is poor, a PDE loss function that embeds the concept of pseudo-time stepping is introduced to avoid this issue. The temporal evolution of the flow field is aligned with the gradient descent process, thereby avoiding the need to solve the unsteady Euler equations. In addition, to improve the accuracy of shock capturing, we propose a simplification method for the steady Euler equations to reduce the difficulty of optimization. By solving four forward problems and one parametric problem, we demonstrate that our method achieves high accuracy in shock capturing and a speedup of approximately 6.7 over an existing advanced method. The key roles of the proposed temporal evolution strategy and Euler equation simplification in achieving accurate shock capturing are confirmed through ablation experiments.

Future work will focus on extending the proposed method to viscous and three-dimensional flows. Particular attention will be paid to shock–boundary-layer interactions and complex three-dimensional shock structures.

## Acknowledgments

This work was supported by the National Natural Science Foundation of China (Grant No. U2441211).

## Appendix A

We demonstrate the effectiveness of using the unsteady Euler equations to constrain NNfoil for shock capturing by solving the inviscid flow around the NACA0012 airfoil at Mach number $Ma = 0.7$ and angle of attack $\alpha = 5°$. NNfoil is used to solve the steady Euler equations and the unsteady Euler equations over the dimensionless time interval $t \in [0, 2]$. It should be noted that here we only demonstrate the capability of NNfoil constrained by the unsteady Euler equations for shock capturing, without performing long-time simulation in the time dimension to obtain the steady state solution satisfying equation (1). The L-BFGS optimizer is employed to perform gradient descent, with the maximum number of inner iterations per epoch set to 1000. The physical domain and the distribution of collocation points for solving the steady Euler equations are the same as those used for Case1 described in Section 3.1 (Fig. 3). The number of collocation points is $N_r = 16080$, corresponding to 80% of all 20100 collocation points, and these points are resampled at each epoch. Boundary conditions $[\rho_\infty, u_\infty, v_\infty, p_\infty] = [1, \cos(\alpha), \sin(\alpha), 1/\gamma Ma^2]$ and $\boldsymbol{V} \cdot \boldsymbol{n} = \boldsymbol{0}$ are imposed on the far-field boundary and the airfoil, respectively, with $N_{bc} = 201$ training points for each. $N_{bc} = 100$ training points are sampled to impose the periodic boundary condition. NNfoil is trained for 20 epochs. $w_{bc} = w_r = 1$.

For the solution of the unsteady Euler equations, considering that the capability of PINN-like methods in solving long time problems is poor [26-28], domain decomposition is employed in the time dimension to improve the accuracy. Specifically, $t \in [0, 2]$ is divided into five subintervals, $t \in [0, 0.4]$, $[0.4, 0.8]$, $[0.8, 1.2]$, $[1.2, 1.6]$, and $[1.6, 2.0]$, and an independent NNfoil model is used for each subinterval, with the subintervals solved sequentially. The initial conditions for the subinterval $t \in [0, 0.4]$ are the uniform fields constructed based on far-field boundary conditions $[\rho_\infty, u_\infty, v_\infty, p_\infty] = [1, \cos(\alpha), \sin(\alpha), 1/\gamma Ma^2]$, while the initial conditions for each remaining subinterval are obtained from the model trained in the previous subinterval. In each subinterval, the space distribution of collocation points is the same as that for the steady equations, and random sampling is performed along the time dimension. The number of collocation points is $N_r = 16080$, and these points are resampled at each epoch. We randomly sample $N_{ic} = 5000$ training points from the collocation points to impose the initial conditions. The boundary conditions for the unsteady problem are the same as those for the steady problem, with $N_{bc} = 2010$ training points sampled on both the far-field boundary and the airfoil, and $N_{bc} = 1000$ training points sampled on the periodic boundary. NNfoil is trained for 60 epochs in each subinterval. $w_{ic} = w_{bc} = w_r = 1$.

Figure A1 illustrates the convergence histories of the loss functions for NNfoil solving the steady and unsteady Euler equations, respectively. We observe that all loss terms have sufficiently converged. Figures A2 and A3 show the pressure coefficient distributions. It can be observed that the results of NNfoil constrained by the steady equations fail to capture the shock, which is consistent with the results in [20]. In contrast, under the constraint of the unsteady equations, NNfoil successfully captures the shock, although slight differences from the reference solution still exist. This result demonstrates the effectiveness of strengthening the physical constraints of NNfoil using the unsteady Euler equations.

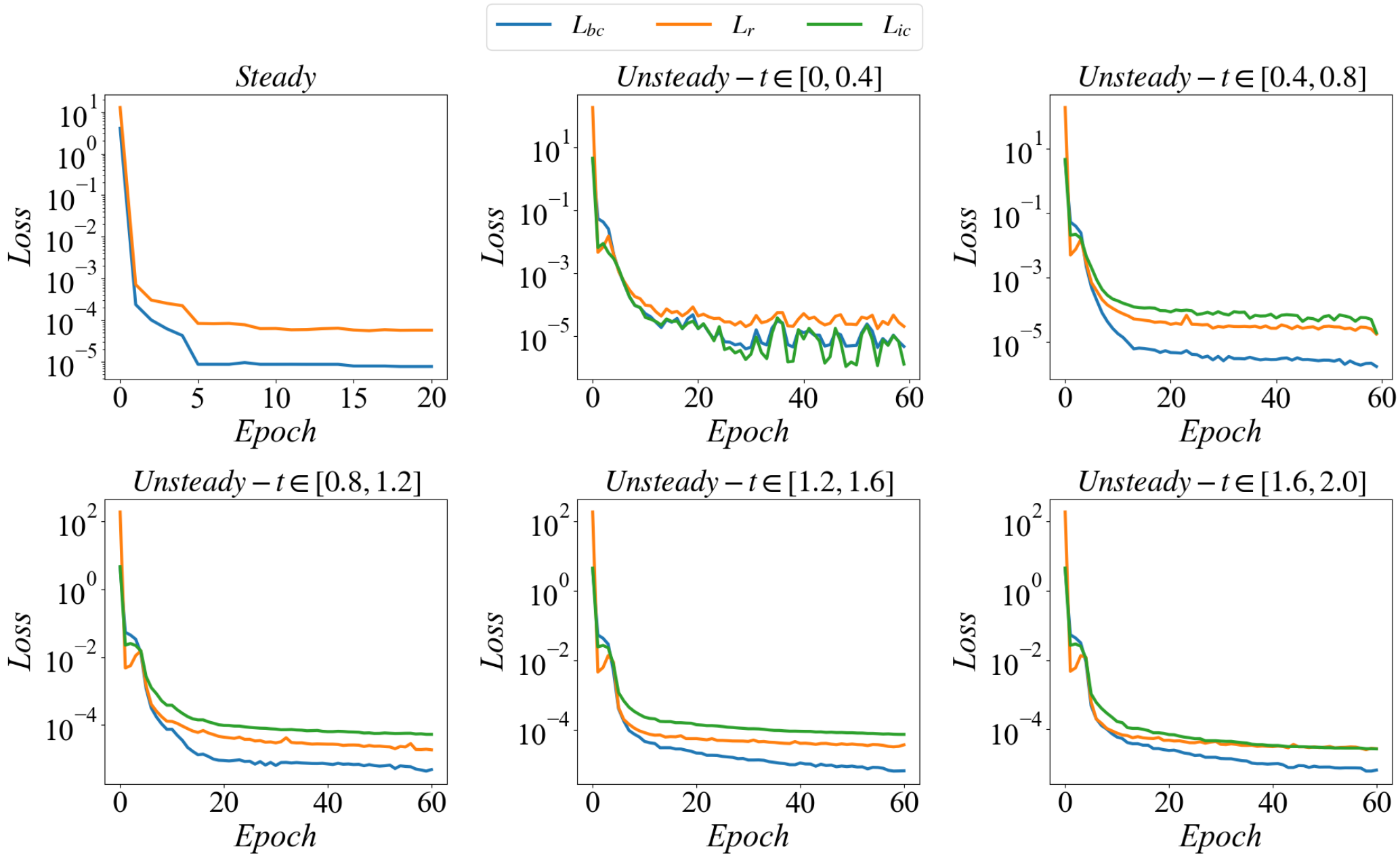

Figure A1. Convergence histories of the loss functions of NNfoil for solving the steady and unsteady Euler equations, respectively.

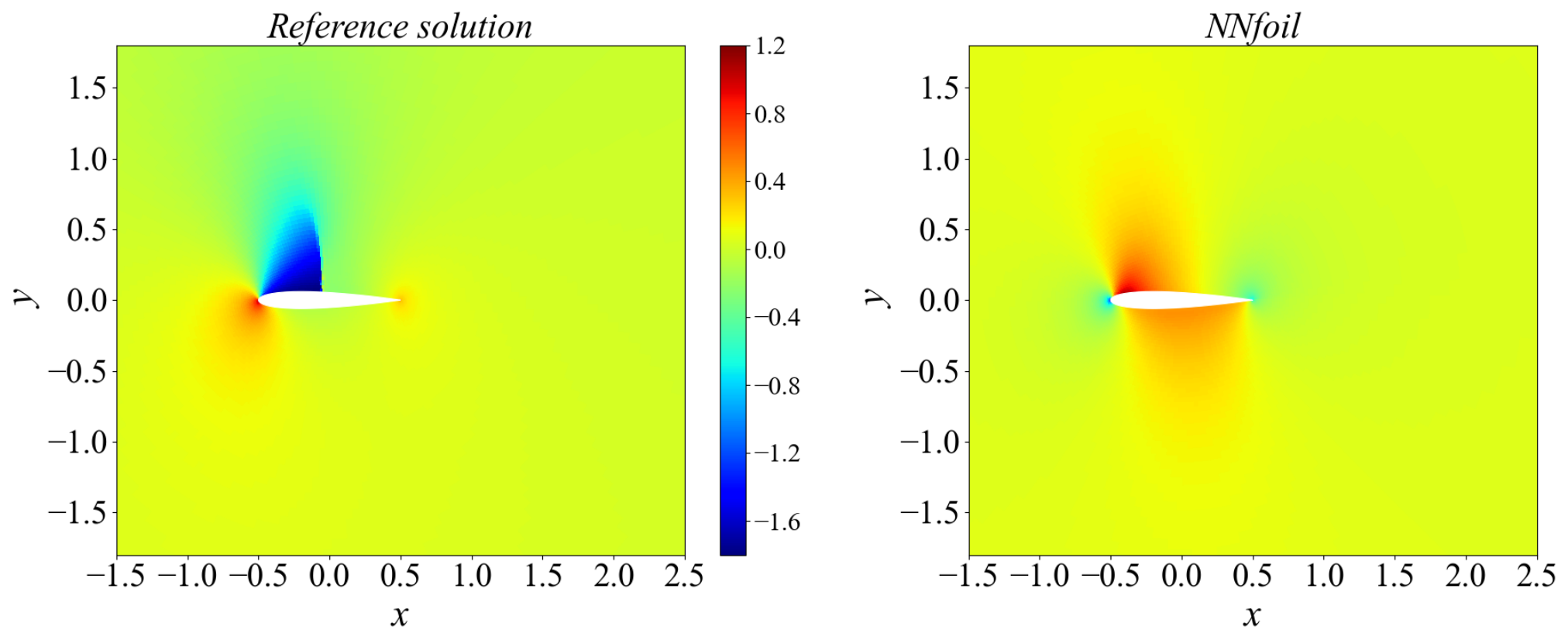

Figure A2. Pressure coefficient distribution obtained by NNfoil for solving the steady Euler equations.

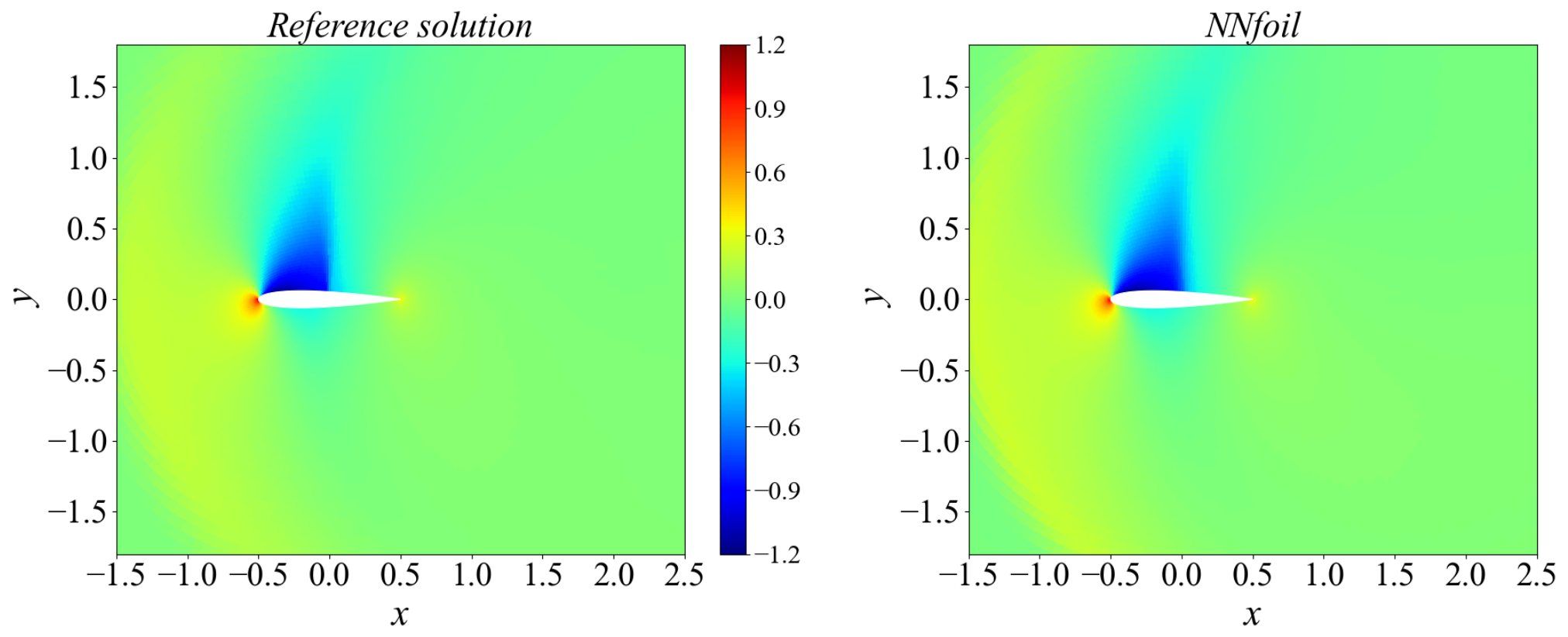

Figure A3. Pressure coefficient distribution obtained by NNfoil at $t = 2.0$ for solving the unsteady Euler equations.

# Appendix B

The method and case labels used in this appendix follow the same conventions as those introduced in Section 3. Figure B1 shows the convergence histories of Algorithm2 using different equation formulations. Algorithm2 uses Eq. (13), while the Algorithm2-Pressure formulation replaces the energy equation in Eq. (13) with Eq. (14). All other settings are identical, as described in Section 3.1. Table B1 summarizes their relative $L_2$ errors in the pressure coefficient distributions of the entire flow field. It can be observed that Algorithm2 generally achieves lower errors than the Algorithm2-Pressure formulation.

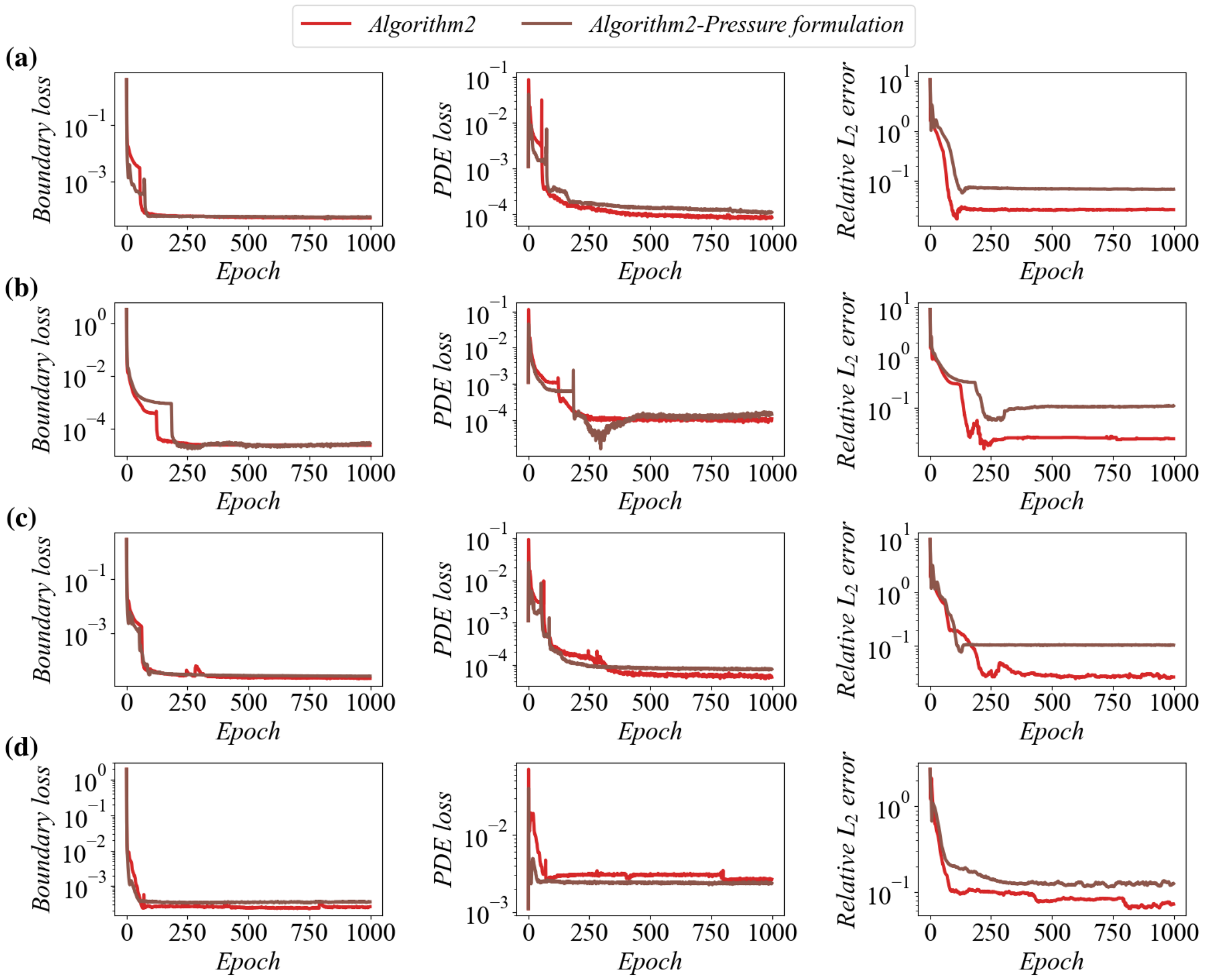

Figure B1. Convergence histories of Algorithm2 using different equation formulations for solving forward problems involving different flow conditions and airfoils. (a) Case1. (b) Case2. (c) Case3. (d) Case4.

# Appendix C

Figures C1–C3 show the pressure coefficient distributions obtained by solving Cases 2–4 in Section 3 using different methods listed in Table 1.

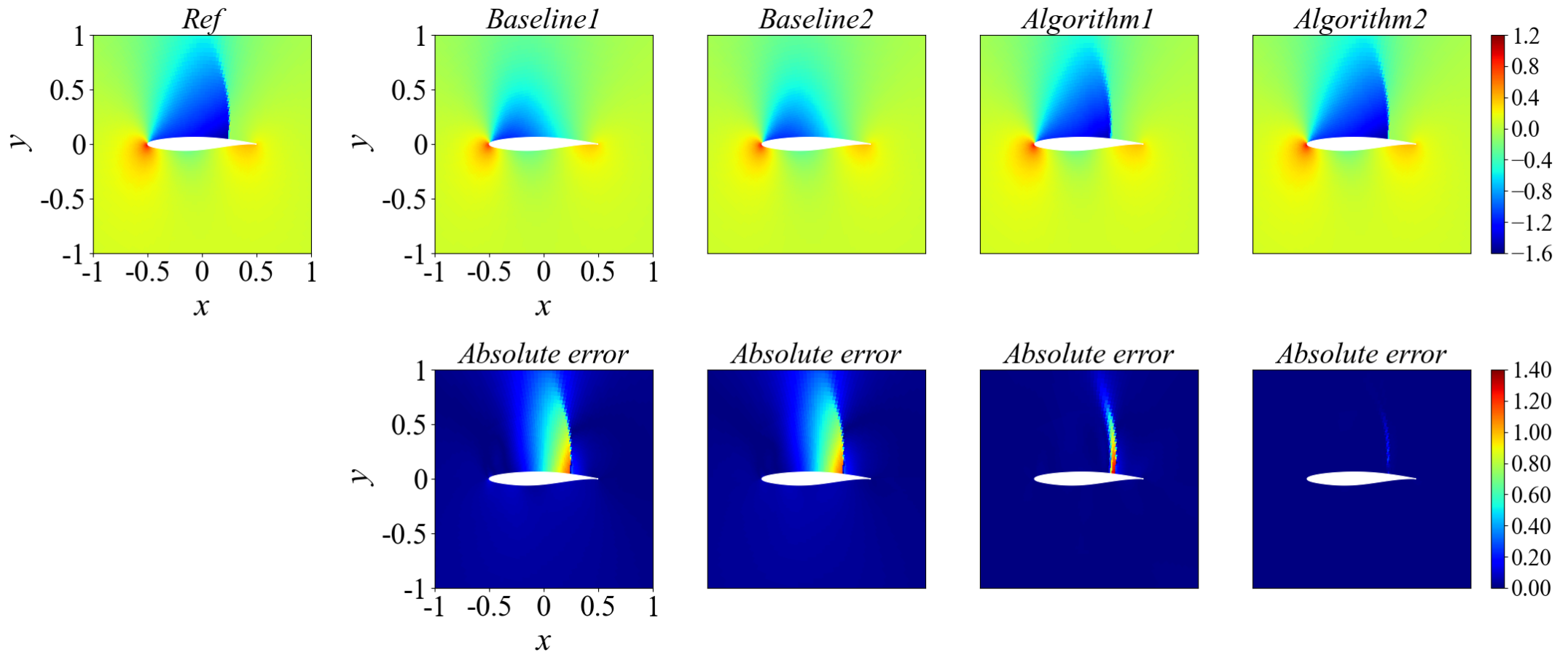

Figure C1. Comparison of pressure coefficient distributions obtained by different methods for Case2.

Table B1. Relative $L_2$ errors (%) in the pressure coefficient distributions of the entire flow field obtained by Algorithm2 and Algorithm2-Pressure formulation.

| | Case1 | Case2 | Case3 | Case4 |
|---|---|---|---|---|
| Algorithm2 | 2.66 | 2.41 | 2.63 | 7.20 |
| Algorithm2-Pressure formulation | 6.78 | 10.85 | 10.40 | 12.54 |

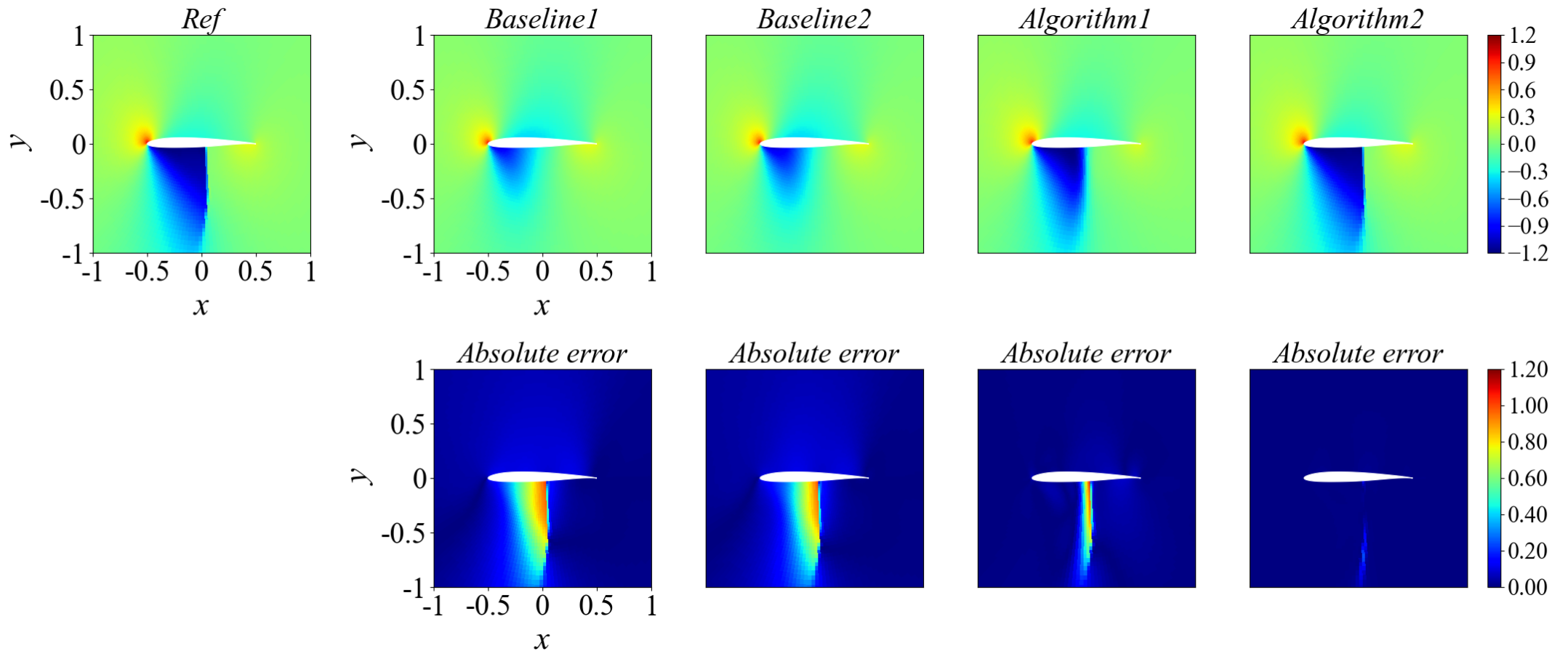

Figure C2. Comparison of pressure coefficient distributions obtained by different methods for Case3.

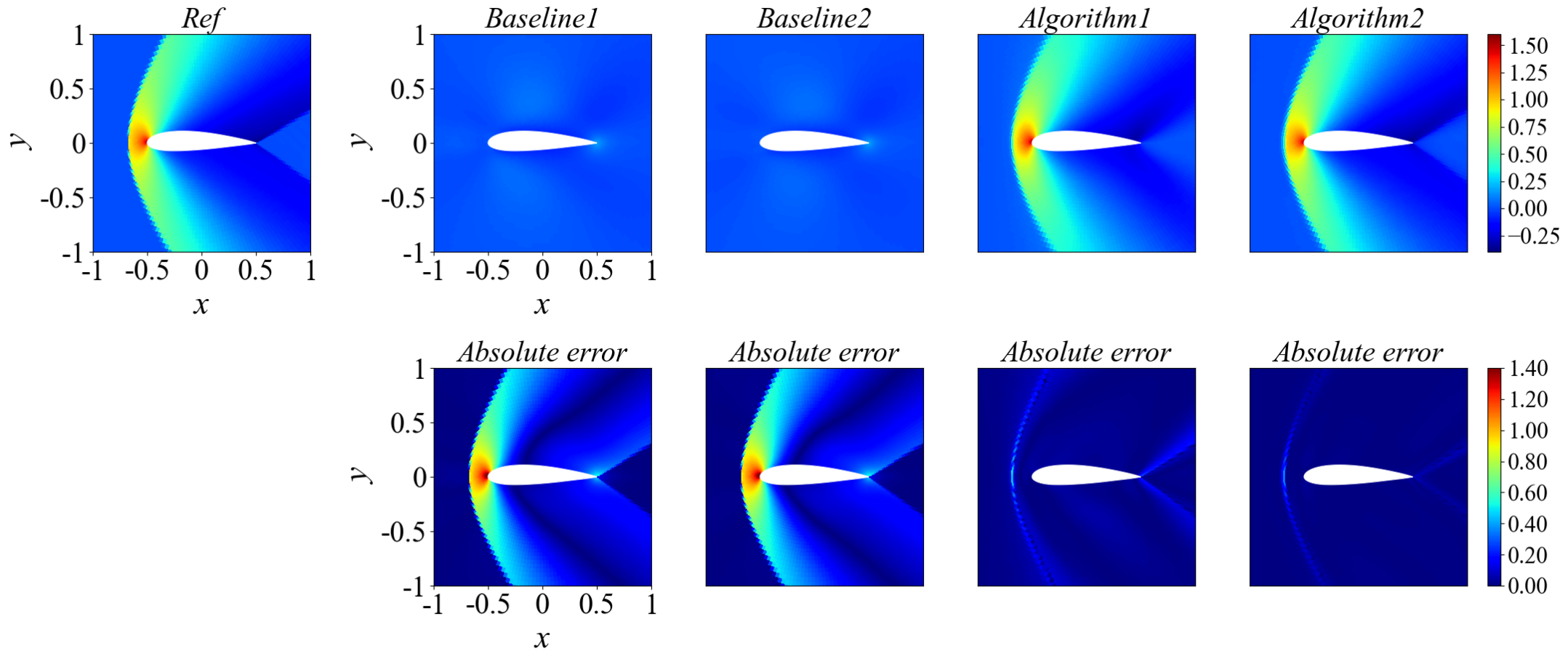

Figure C3. Comparison of pressure coefficient distributions obtained by different methods for Case4.

# Appendix D

Figure D1 shows the convergence histories of the four methods for Case1, where the curves represent the mean and the shaded regions indicate two standard deviations over five runs with different random seeds. We observe that all methods are robust.

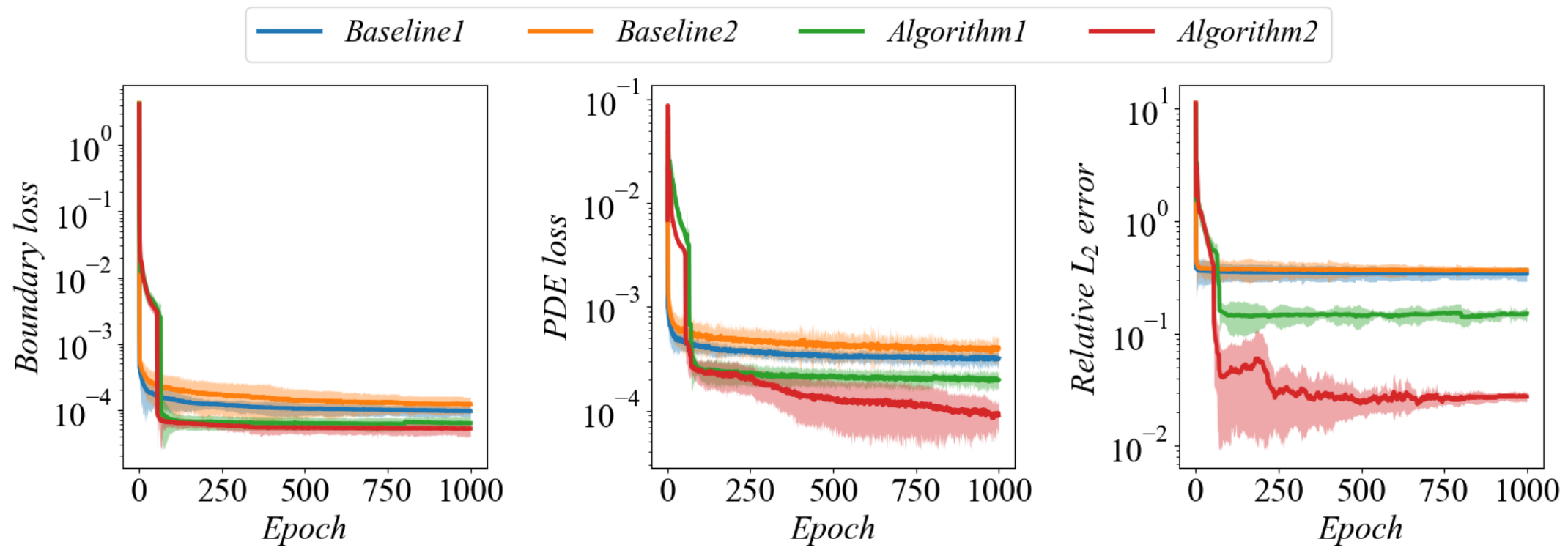

Figure D1. Convergence histories of the four methods for solving Case1. The curves and shaded regions represent the mean and two standard deviations.